\documentclass{aa}

\usepackage[varg]{txfonts}
\usepackage{graphicx}
\usepackage{natbib}
\usepackage{siunitx}
\usepackage{mathrsfs, amsmath, adjustbox}
\usepackage{flushend}
\usepackage[hypertexnames=false]{hyperref}
\hypersetup{colorlinks=true, citecolor=blue, linkcolor=blue}

\bibpunct{(}{)}{;}{a}{}{,}

\newcommand{\HL}{\mathscr{L}}
\newcommand{\bv}{\boldsymbol{v}}
\newcommand{\br}{\boldsymbol{r}}
\newcommand{\bk}{\boldsymbol{k}}
\newcommand{\bb}{\boldsymbol{B}}

\newcommand{\beb}{\boldsymbol{e_B}}
\newcommand{\ba}{\boldsymbol{A}}
\newcommand{\unittensor}{\boldsymbol{I}}
\newcommand{\kappatensor}{\boldsymbol{\kappa}}

\begin{document}

\title{Thermal instabilities: Fragmentation and field misalignment of filament fine structure}

\author{N. Claes\inst{1} \and R. Keppens\inst{1} \and C. Xia\inst{2}} 

\institute{Centre for mathematical Plasma-Astrophysics, Celestijnenlaan 200B, 3001 Leuven, KU Leuven, Belgium
        \and
        Yunnan University, Kunming, PR China \\
        \email{niels.claes@kuleuven.be}; \email{rony.keppens@kuleuven.be}; \email{chun.xia@ynu.edu.cn}}

\date{Accepted 14 March 2020}

\let\oldpageref\pageref
\renewcommand{\pageref}{\oldpageref*}


\abstract
{Prominences show a surprising amount of fine structure and it is widely believed that their threads, as seen in H$\alpha$ observations, provide indirect information concerning magnetic field topology. Both prominence and coronal rain condensations most likely originate from thermal instabilities in the solar corona. It is still not understood how non-linear instability evolution shapes the observed fine structure of prominences. Investigating this requires multidimensional, high-resolution simulations to resolve all emerging substructure in great detail.}
{We investigate the spontaneous emergence and evolution of fine structure in high-density condensations formed through the process of thermal instability under typical solar coronal conditions. Our study reveals intricate multidimensional processes that occur through in situ condensations in a representative coronal volume in a low plasma beta regime.}
{We quantified slow wave eigenfunctions used as perturbations and discuss under which conditions the thermal mode is unstable when anisotropic thermal conduction effects are included. We performed 2D and 3D numerical simulations of interacting slow magnetohydrodynamic (MHD) wave modes when all relevant non-adiabatic effects are included. Multiple levels of adaptive mesh refinement achieve a high resolution near regions with high density, thereby resolving any emerging fine structure automatically. Our study employs a local periodic coronal region traversed by damped slow waves inspired by the presence of such waves observed in actual coronal magnetic structures.}
{We show that the interaction of multiple slow MHD wave modes in a regime unstable to the thermal mode leads to thermal instability. This initially forms pancake-like structures almost orthogonal to the local magnetic field, 
while low-pressure induced inflows of matter generate rebound shocks. This is succeeded by the rapid disruption of these pancake sheets through thin-shell instabilities evolving naturally from minute ram pressure imbalances. This eventually creates high-density blobs accompanied by thread-like features from shear flow effects. The further evolution of the blobs follows the magnetic field lines, such that a dynamical realignment with the background magnetic field appears.
However, the emerging thread-like features are not at all field-aligned, implying only a very weak link between fine structure orientation and magnetic field topology.}
{As seen in our synthetic H$\alpha$ views, threads formed by non-linear thermal instability evolution do not strictly outline magnetic field structure and this finding has far-reaching implications for field topology interpretations based on H$\alpha$ observations.}

\maketitle


\section{Introduction}
Solar prominences are large, cool, and dense structures extending thousands of kilometres outwards from the solar surface into the hot corona. Thermal instabilities lie at the basis behind the formation of these intriguing and complex structures, which is a theory that dates back to seminal papers by
\citet{parker1953} and \citet{field1965}. These works explained how a runaway radiative cooling effect can originate from thermal instability (TI) in plasmas. It was soon realised \citep{smith1977, priest1979} that these kind of instabilities, when occurring in solar coronal conditions, could be responsible for the formation of condensations that are orders of magnitude denser and colder than the initial thermal equilibrium conditions.
The characteristic runaway radiative cooling process driving the formation of these structures rapidly becomes a non-linear process, such that high-resolution numerical simulations are required to study their temporal evolution and underlying substructure in great detail.

The presence of TI as a driving process behind prominence formation has been confirmed using direct observations of the solar corona. \citet{berger2012} showed through SDO/AIA observations that a quiescent prominence forms via a condensation
process of hot plasma in a coronal cavity, which subsequently cools down from 2 MK to about 0.08 MK,  consistent with a runaway TI process. The same holds true for observations of coronal rain, showing plasma evaporation and condensation followed by the formation of high-density blobs that eventually fall back down onto the solar surface, where the shape and intensity of this phenomenon is further dependent on loop geometry and heating contributions \citep{auchere2018}.
Many observations in recent decades revealed the omnipresent appearance of coronal rain \citep{degroof2005,kamio2011, antolin2011, antolin2012, antolin2015} and fine, elongated threads of plasma pervading solar prominences \citep{engvold1998, lin2005, ballester2006, mackay2010}.
However, there is ambivalence between various findings, some observations indicate that a prominence consists of multiple horizontal threads \citep{casini2003, okamoto2007}, while others reported the presence of various vertical threads \citep{berger2008, chae2008}. The general consensus however is that threads as seen in H$\alpha$ observations yield some information about the topology of the magnetic field. The actual origin of this prominence fine structure is at present still unknown, although some early analytical work based on linear MHD theory supports the hypothesis that the anisotropic thermal conduction (in particular, inclusion of finite, but small perpendicular conduction) can induce fine structure in the unstable linear eigenmodes \citep{VDL1991_2}.

The typical low coronal plasma beta conditions are then frequently invoked to argue for a convenient dimensional reduction, in which the prominence or coronal rain plasma is forced to live and evolve on given rigid field line shapes and the downwards gravitational pull on the plasma implies that it can only be collected in dipped field line portions. Other works completely ignore the actual plasma and instead focus on the magnetic skeleton (for a review, see \citet{parenti2014, gibson2018}), ensuring a force-free magnetic configuration in which dipped field line portions are interpreted as sites for prominence matter. These two approaches have been unified in \citet{luna2012}, in which many 1D hydrodynamic simulations were presented as a realistic 3D prominence model. Similar combinations of 3D force-free magnetic models, with hydrostatic equilibrium along field lines, could be turned into whole prominence fine structure models \citep{gunar2015, gunar2016}, emphasising the synthetic emissions from these models.

If indeed all chromospheric fibrils in filament channels and actual filament threads give us magnetic field topology, 3D non-adiabatic MHD simulations can be avoided, and we fall back to hydrodynamic problems in 1D along fixed field line shapes. Three decades of research following this approach have been very successful in providing insights into the physics of condensations, either for filaments or coronal rain, in a variety of given field line topologies \citep{antiochos1999, mok1990, antiochos2000, karpen2001, luna2012, xia2011}. These models have always looked at thermodynamically and gravitationally structured loops, and continue to raise debates on the role of so-called thermal non-equilibrium (TNE) versus TI  \citep{klimchuk2019, antolin2019}.

Extensions of the 1D models including chromospheric evaporation to start up either TNE cycles or form a condensation by TI are urgently needed to address realistic coronal rain and prominence formation scenarios, the role of (condensation-induced) magnetic dips, and all aspects that deviate from perfect field alignment. The first 2.5D simulation of the formation of a quiescent prominence by chromospheric heating was done by \citet{xia2012}. These authors showed how plasma evaporation followed by condensation due to TI captures all phases of the formation process and that the magnetic field must be deformed to support heavy prominence plasma even in low beta conditions. The simulated filament structure formed its own magnetic dips and realised force-balance in a magnetic arcade system. Evaporation-condensation was further demonstrated in quadrupolar field arcades yielding funnel prominences \citep{keppens2014}, while coronal rain was studied in 2D \citep{fang2013, fang2015shocks} or 3D configurations \citep{moschou2015, xia2017}. In 3D flux ropes, thermodynamically consistent models for a solar prominence were initiated in \citet{xia2014} and evolved to demonstrate a clear formation cycle of fine-structured condensations \citep{xia2016}. The latter showed a continuous reappearance of threads and blobs, moving downwards along mass-loaded field lines, complicated by shear flow effects. Synthetic extreme ultraviolet (EUV) images resemble many aspects of true prominence observations. Eruptive prominences have been modelled by \citet{fan2017}, as yet representing the prominence body by a single, fairly homogeneous large-scale condensation. The same holds true for models from \citet{kaneko2015, kaneko2017}. Recent work by \citet{li2018} has shown that magnetic reconnection events between solar loops may play a complementary role in triggering TI, although in this work we only focus on slow MHD waves pervading a region of plasma with conditions comparable to a local coronal volume.

Motivated by these recent developments in multidimensional prominence and rain modelling and determined to focus on the fine structure aspect, we look at the temporal evolutions of condensations formed through the process of TI, in the presence of anisotropic thermal conduction. To dismiss all possible arguments concerning the need to distinguish TNE from TI processes, we ignore the thermodynamic and gravitational structure in the large-scale configuration, and isolate a local coronal volume that is pervaded by waves. Slow waves have been detected in various coronal structures~\citep{king2003, roberts2005, demoortel2015} and we simply take a uniform low plasma beta box as a background, ignoring gravity. In this scenario, the only process behind condensation formation is TI.

The main motivation behind this is threefold: first, it has been shown that the solar corona shows ubiquitous presence of slow MHD wave modes, which naturally interact with each other. Secondly, a homogeneous periodic box with solar coronal conditions can serve as an approximation of a local part of the corona, lightly perturbed by passing slow waves. Finally, such a back-to-basics approach allows us to omit additional physical effects such as gravity or evaporation-condensation processes and associated flow patterns, which further complicate detailed analysis. This extends our previous work with analogous local 2D simulations \citep{claes2019} by including anisotropic thermal conduction, three-dimensional simulations, and much higher resolutions.

In Section~\ref{sect: equations} we lay the mathematical groundwork by carrying out a Fourier analysis on the MHD equations and calculating slow wave eigenfunctions, which are used as perturbations when setting up the simulations in Section~\ref{sect: numerical_setup}. Since this approach also requires knowledge of the eigenfrequency $\omega$, stability regimes of the thermal and slow modes in solar coronal conditions are calculated in 
Section~\ref{subsect: thermal_mode}. Using the full mathematical prescription for the eigenfunctions together with the corresponding eigenvalues $\omega$ ensures a consistent numerical excitation of slow MHD wave modes, done in a homogeneous, Cartesian box with periodic boundary conditions. The emerging fine structure is studied in detail for both 2D and 3D simulations in Section~\ref{sect: results}, where the latter are supplemented by detailed volume renderings and synthetic H$\alpha$ views in Section~\ref{subsect: 3D}.

\section{Governing equations}           \label{sect: equations}
\subsection{Linearisation}
The magnetohydrodynamic (MHD) equations with the addition of conduction effects and a radiative cooling contribution can be written as
\begingroup
\allowdisplaybreaks
\begin{align}
        \frac{\partial \rho}{\partial t} &+ \nabla \cdot (\rho\bv) = 0,                                                 \label{eq: cons_continuity}        \\
        \rho\frac{\partial \bv}{\partial t} &+ \rho \bv \cdot \nabla\bv + \nabla p - (\nabla \times \bb) \times \bb = \boldsymbol{0},      \label{eq: cons_momentum}\\
        \rho\frac{\partial T}{\partial t} &+ \rho \bv \cdot\nabla T + (\gamma - 1)p\nabla \cdot \bv + (\gamma - 1)\rho \HL \notag \\
                                                                          &- (\gamma - 1) \nabla \cdot (\kappatensor \cdot \nabla T) = 0,        \label{eq: cons_energy} \\
        \frac{\partial \bb}{\partial t} &- \nabla \times (\bv \times \bb) = \boldsymbol{0,}                                        \label{eq: cons_induction}
\end{align}
\endgroup
where the quantities $\rho$, $\bv$, $T,$ and $\bb$ denote the density, velocity, temperature, and magnetic field, the latter given in units where $\mu_0 = 1$. The ratio of specific heats is denoted by $\gamma$ and is equal to $5/3$.
The symbol $\HL$ indicates the heat-loss function, defined as the energy losses minus the energy gains due to radiative cooling effects \citep{field1965}. The system of equations is closed by the ideal gas law
\begin{equation}
        p\mu = \mathscr{R}\rho T,
\end{equation}
in which $\mathscr{R}$, $\mu,$ and $p$ represent the gas constant, mean molecular weight, and pressure, respectively. Because thermal conduction in the presence of a magnetic field is mostly aligned with the field lines, we use a tensor representation to model this anisotropy. The thermal conductivity tensor $\kappatensor$ is defined as
\begin{equation}                \label{eq: conduction}
        \kappatensor = \kappa_\parallel \beb\beb + \kappa_\bot(\unittensor - \beb\beb),
\end{equation}
where $\unittensor$ denotes the unit tensor and $\beb = \bb / B$ is a unit vector along the magnetic field, and the quantities $\kappa_\parallel$ and $\kappa_\bot$ indicate the thermal conduction coefficients parallel and perpendicular to the magnetic field lines, respectively. We typically take the Spitzer conductivity for $\kappa_\parallel$ in astrophysical applications \citep{spitzer2006}, defined as
\begin{equation}
        \kappa_\parallel = 8 \times 10^{-7} T^{5/2}     ~~~ \text{erg cm$^{-1}$ s$^{-1}$ K$^{-1}$},
\end{equation} 
while the perpendicular conduction coefficient is given by
\begin{equation}
        \kappa_\bot = 4 \times 10^{-10} n^2 B^{-2} T^{-3} \kappa_\parallel,
\end{equation}
where $n$ is the number density. In typical solar coronal condictions $\kappa_\bot \approx 10^{-12}\kappa_\parallel$ \citep{braginskii}, such that $\kappa_\parallel$ is dominant and thermal conduction is almost unidirectional along the magnetic field lines.

Next we linearise the MHD equations assuming an infinite, static, and homogeneous background denoted by a subscript $0$ where $\bv_0 = 0$. Small deviations from the equilibrium state are denoted by a subscript $1$, leading to the following system of linearised equations
\begin{align}
        \frac{\partial \rho_1}{\partial t} &+ \nabla \cdot (\rho_0 \bv_1) = 0,            \label{eq: continuity}  \\
        \rho_0\frac{\partial \bv_1}{\partial t}  &+  \nabla(T_0\rho_1 + \rho_0T_1) - (\nabla \times \bb_0) \times (\nabla \times \ba_1)    \label{eq: momentum}    \\
                                                                                                                                        &- \left[\nabla \times (\nabla \times \ba_1)\right] \times \bb_0 = \boldsymbol{0},                 \notag          \\
        \rho_0\frac{\partial T_1}{\partial t} &+ (\gamma - 1)\rho_0 T_0 \nabla \cdot \bv_1 +(\gamma - 1)\rho_1 \HL_0   \label{eq: energy}      \\
                                                                                                                  &+ (\gamma - 1)\rho_0(\HL_T T_1 + \HL_\rho \rho_1) - (\gamma - 1)\nabla \cdot (\kappatensor_0 \cdot \nabla T_1) = 0,            \notag  \\
        \frac{\partial \ba_1}{\partial t} &- \bv_1 \times \bb_0 = \boldsymbol{0}.               \label{eq: induction}
\end{align}
The background dimensionless pressure $p_0$ is written as $\rho_0 T_0$, while the perturbed pressure contribution $p_1$ is substituted by $T_0\rho_1 + \rho_0T_1$ following the linearised ideal gas law. The temperature and density derivatives of the heat-loss function are denoted by
$\HL_T$ and $\HL_\rho$, respectively. The perturbed magnetic field is rewritten in terms of a vector potential $\bb_1 = \nabla \times \ba_1$, such that the divergence free condition on $\bb_1$ is naturally satisfied.

Next we perform a standard Fourier analysis on the perturbed quantities by assuming plane-wave perturbations of the form 
\begin{equation}        \label{eq: Fourier}
        \rho_1 = \hat{\rho}_1 \exp(i \bk \cdot \br - i\omega t),
\end{equation}
and the same for all other variables. This transforms the partial differential equations \eqref{eq: continuity}-\eqref{eq: induction} into a system of algebraic equations, given by
\begin{align}
        \omega \rho_1 &= \rho_0 \bk \cdot \bv_1,                        \label{eq: rho1}   \\
        \omega \bv_1  &= \frac{T_0}{\rho_0}\bk\rho_1 + \bk T_1 - \frac{i}{\rho_0}\left[\bk \times (\bk \times \ba_1)\right] \times \bb_0,          \\
        \omega T_1     &= (\gamma - 1)T_0 \bk \cdot \bv_1 - i(\gamma - 1)(\HL_T T_1 + \HL_\rho \rho_1)           \\
                                                         &- i(\gamma - 1)\frac{1}{\rho_0}(\kappa_\parallel k_\parallel^2 + \kappa_\bot k_\bot^2)T_1,       \notag  \\
        \omega \ba_1  &= i\bv_1 \times \bb_0.           \label{eq: a1}
\end{align}
As confusion is not possible we omitted the hat notation used in Eq. \eqref{eq: Fourier}. The eight equations given above form a complete eigenvalue problem, with the eigenvectors containing the variables $\rho_1$, $\bv_1$, $T_1$, and $\ba_1$, such that the eigenfrequency $\omega$ of each respective mode can be obtained by directly solving for the eigenvalues of the corresponding matrix. The components of the wave vector $\bk$ parallel and perpendicular to $\bb_0$ are denoted by $k_\parallel$ and $k_\bot$, respectively.

\subsection{Eigenfunctions}
In order to consistently excite the MHD wave modes associated with a certain eigenvalue $\omega$ in numerical simulations, we calculated the exact eigenfunctions for a given orientation of the wave vector $\bk$ with respect to the magnetic field $\bb_0$. The usual convention is to align the magnetic field itself with one of the coordinate axes; in this work we do the opposite and align $\bk$ instead. The advantage of this is that we can use periodic boundary conditions on all sides of the domain in our numerical simulations when exciting the eigenfunctions, even when imposing multiple wave vectors.
For $\bk$ aligned with the $x$-axis in three dimensions we defined
\begin{equation}
        \bk = (k_x, ~0, ~0),            \qquad          \bb_0 = (B_{0x}, ~B_{0y}, ~B_{0z}),
\end{equation}
such that the eigenfunctions for this case are given by 
\begin{align}
        \rho_1 &= \frac{k_x \rho_0 }{\omega} v_x,               \label{eq: ef_rho} \\
        v_y     &= \frac{B_{0x} B_{0y} k_x^2}{B_{0x}^2 k_x^2 - \omega^2 \rho_0} v_x,                            \\
        v_z             &= \frac{B_{0x} B_{0z} k_x^2 }{B_{0x}^2 k_x^2 - \omega^2 \rho_0} v_x,                            \\
        T_1             &= - \frac{k_x \rho_0 (\gamma - 1)\left(i L_\rho\rho_0 - T_0 \omega\right)}{\omega \left[i (\gamma - 1) K + i (\gamma - 1)L_T\rho_0 + \omega \rho_0\right]} v_x,            \\
        A_y             &= \frac{i B_{0z} \omega \rho_0 }{B_{0x}^2 k_x^2 - \omega^2 \rho_0} v_x,                 \\
        A_z             &= - \frac{i B_{0y} \omega \rho_0 }{B_{0x}^2 k_x^2 - \omega^2 \rho_0} v_x,         \label{eq: ef_az}
\end{align}
where the eigenfunction for the $x$-component of the vector potential $A_x = 0$, and we defined $K = \kappa_\parallel k_\parallel^2 + \kappa_\bot k_\bot^2$ to abbreviate the notation. All of the above eigenfunctions have a dependence on $v_x$, such that this eigenfunction can be chosen arbitrarily. 
Similar results can be obtained when $\bk$ is aligned along a different coordinate axis or when shifting to 2D. The analysis above is standard procedure, but generalises our earlier paper \citep{claes2019} in which we did not yet quantify the role of finite anisotropic thermal conduction on the eigenfunctions. These eigenfunction quantifications augment the original works by \citet{parker1953} and \citet{field1965}, where only the dispersion relation was discussed in great detail.

\section{Numerical set-up}       \label{sect: numerical_setup}
We made use of the parallelised Adaptive Mesh Refinement Versatile Advection Code MPI-AMRVAC \citep{amrvacpaper1, porth2014, amrvacpaper2} to numerically solve the full non-linear, non-adiabatic MHD equations on a two- or three-dimensional Cartesian mesh with multiple levels of adaptive mesh refinement (AMR).
The base resolution used in all our simulations is $100^2$ and $100^3$ for the 2D and 3D results, respectively. Three (3D) and five (2D) levels of AMR were used with the trigger of refinement based on a density criterion, achieving a maximum effective resolution of $400^3$ and $1600^2$ near high-density regions in 3D and 2D, respectively. As indicated further, one ultra-high-resolution run in 2D uses seven AMR levels for a $6400^2$ run to reveal more details. The length scale of the system is of the order of a few to $10$ Mm depending on the simulation. We considered a fully ionised plasma with a $0.1$ helium abundance such that $\rho = 1.4 m_p n_H$, where $m_p$ denotes the proton mass and
$n_H$ the hydrogen number density. The ideal gas law for such a mixture is then written as $p = 2.3k_B n_H T$, where $k_B$ is the Boltzmann constant.

Both the radiative cooling contribution and conduction effects are handled as an additional source term to the energy equation as shown in Eq.~\eqref{eq: cons_energy}. The radiative cooling term is defined as $\rho \HL = 1.2 n_H^2 \Lambda(T) - H_0$, where $\Lambda(T)$ denotes the radiative cooling curve assuming optically thin radiative losses and $H_0$ the heating term. For $\Lambda(T)$ we adopted the cooling curve defined by \citet{SPEXcurve}, augmented at low temperatures (below $\approx$ 10 kK) by that from \citet{dalgarno1972}. The heating term is taken to be constant over time and equal to $1.2 n_H^2 \Lambda(T)$ at $t=0$. This implies that the radiative losses and heating contribution balance each other out at $t=0$ such that an initial state of thermal equilibrium is achieved. Thermal conduction is added in full tensorial form as defined by Eq.~\eqref{eq: conduction}. A detailed explanation on how the code handles conduction effects parallel and perpendicular to the magnetic field is provided in \citet{amrvacpaper2}.

\subsection{Initial conditions}
We used periodic boundary conditions on all sides of the domain and started from a uniform thermal equilibrium state as described before. The wave vector $\bk$ is always equal to $2\pi$ divided by the domain length to ensure one single oscillation along one of the axes to satisfy the periodic boundary conditions on all sides. The magnetic field orientation with respect to the coordinate axes is described by an angle $\theta$, measured from the $xy$-plane towards the $z$-axis, and by an angle $\delta$ measured in the $xy$-plane from the $x$-axis towards the $y$-axis. Using this description the background magnetic field $\bb_0$ can be written as
\begin{equation}        \label{eq: B0}
        \bb_0 = (B_0 \cos\theta \cos\delta, ~B_0 \cos\theta \sin\delta, ~B_0 \sin\theta).
\end{equation}
In two dimensions the expression for $\bb_0$ becomes straightforward, as we only have a single angle $\theta$ between $\bb_0$ and $\bk$. Because of possible ambiguity in three dimensions when talking about wave vector components `perpendicular' to the magnetic field lines we defined these as
\begin{equation}
        k_{x, \parallel} = \frac{\cos\theta}{\cos\delta}k_x             \qquad\text{and}\qquad                 k_{x, \bot} = \frac{\sin\theta}{\cos\delta}k_x,
\end{equation}
for a wave vector $\bk$ aligned with the $x$-axis. For the initial perturbations we used the eigenfunctions as given in Eqs. \eqref{eq: ef_rho}-\eqref{eq: ef_az}, where the eigenfrequency $\omega$ is obtained by numerically solving the eigenvalue problem $A\boldsymbol{x}  = \omega\boldsymbol{x}$ using the Python \texttt{SciPy} package, where $\boldsymbol{x}$ is the vector $(\rho_1, \bv_1, T_1, \ba_1)$ and $A$ the $8\times8$ matrix containing the corresponding coefficients as given by Eqs. \eqref{eq: rho1}-\eqref{eq: a1}. For simulations without conduction effects the parameter $K$ can be set equal to zero.

The obtained eigenvalues are complex owing to the inclusion of non-adiabatic effects. The eigenvalue problem yields seven eigenvalues, two of which are real and correspond to a forwards and backwards travelling Alfv\'en wave. These waves are not affected by non-adiabatic effects in a homogeneous medium \citep{field1965}, so we do not consider these waves further. One eigenvalue is purely imaginary and corresponds to the thermal mode, which is responsible for TI. The final four eigenvalues are two times two complex conjugates, corresponding to slow and fast MHD waves. The real part of the complex eigenvalue
corresponds to the oscillation frequency of the wave as a result of the Fourier analysis we performed in Eq. \eqref{eq: Fourier}, while the complex part indicates whether a wave is damped or unstable. Instability corresponds to a positive imaginary part $\omega_I$, while stability
originates from a negative $\omega_I$.

All simulations were initiated using slow MHD waves, that is, taking the eigenvalue corresponding to the slow mode solution and using that to initialise the eigenfunctions. For the initial perturbation $v_x$ we took a plane wave given by
\begin{equation}        \label{eq: perturbation}
        v_x = C\left[\cos(k_x x) + i\sin(k_x x)\right],
\end{equation}
where $C$ is some constant amplitude, taken to be $10^{-3}$. Similar expressions are used when $\bk$ is aligned along a different axis, for example along the $y$-axis, all eigenfunctions become dependent on $v_y$, which can again be taken arbitrarily and similar to Eq. \eqref{eq: perturbation}. When initialising the simulations we take the real part of the complex eigenfunctions.

Unless explicitly stated otherwise, all equilibrium configurations considered have a density of $\rho_n = 2.34 \times 10^{-15}$ g cm$^{-3}$ (a standard coronal value) and a magnetic field strength of $10$ gauss. This field value is typical for prominences \citep{gibson2018}. The pressure in all simulations is calculated using the ideal gas law,
following $p \sim \rho T$ in normalised code units. The combination of periodic boundary conditions on all sides and a conservative numerical scheme ensures that the total mass across the domain is conserved at all times, such that any condensation is thus an in situ process where mass gets redistributed but not created or destroyed (i.e. unlike in evaporation-condensation configurations, where mass accumulates in the corona from evaporating chromospheric matter).

\subsection{Magnetic field treatment and div(B)}
When doing calculations in MHD we can always make sure that the divergence-free condition on the magnetic field $\nabla \cdot \bb = 0$ is satisfied analytically. However, it is not so trivial to fulfil this requirement in simulations and different methods are available to tackle this problem. One possibility is adding sources proportional to numerical monopole errors or doing parabolic cleaning to diffuse any local error compliant with the limits on the time step. In our simulations we used the newly developed \texttt{constrained transport} feature of MPI-AMRVAC, which was first implemented and demonstrated in its general relativistic MHD code variant BHAC \citep{porth2017bhac}.
This method ensures that if the initial magnetic field, staggered on cell face centres, is integrated from a vector potential on the cell edges then the initial magnetic field divergence at cell centres is zero up to machine precision in a particular discretisation. Hence, we initialised the vector potential on cell edges for the initial equilibrium state, as is implemented in our code.
We can therefore define $\ba_0$ for the equilibrium state as
\begin{equation}
        \ba_0 = (B_{0y}z, ~B_{0z}x, ~B_{0x}y).
\end{equation}
This retrieves the original homogeneous $\bb_0$ when calculating $\bb_0 = \nabla \times \ba_0$, and the real part of the perturbations (eigenfunctions) in the vector potential are consequently added to the above expressions.

\section{Results}       \label{sect: results}
\subsection{Thermal mode instability}   \label{subsect: thermal_mode}
Before fixing all equilibrium conditions (in essence, choosing a temperature value and a particular magnetic field orientation), we have to pay attention to the actual stability and instability regimes of the thermal mode in the presence of thermal conduction.
The latter always has a stabilising effect on TI because temperature variations are rapidly smoothed out by field-aligned conduction. However, we can always quantify the critical wave number $k_\text{crit}$ \citep{field1965}. In essence, perturbations with a wave number above $k_\text{crit}$ (or with a small enough wavelength) are completely stabilised by thermal conduction. By changing the length scale of our numerical domain, we effectively change the wave number in the direction that the perturbation is applied, while still keeping a single oscillation to satisfy the periodic boundary conditions. In taking a longer wavelength, we are able to excite an unstable thermal mode even with the inclusion of thermal conduction. Thermal instability itself is not triggered if the homogeneous equilibrium configuration is stable to the thermal mode, since in this case all MHD wave modes are damped, implying that the slow-wave perturbations we model will dampen out and no runaway cooling will ever occur.
\begin{figure}[h]
        \centering
        \includegraphics[width=\columnwidth]{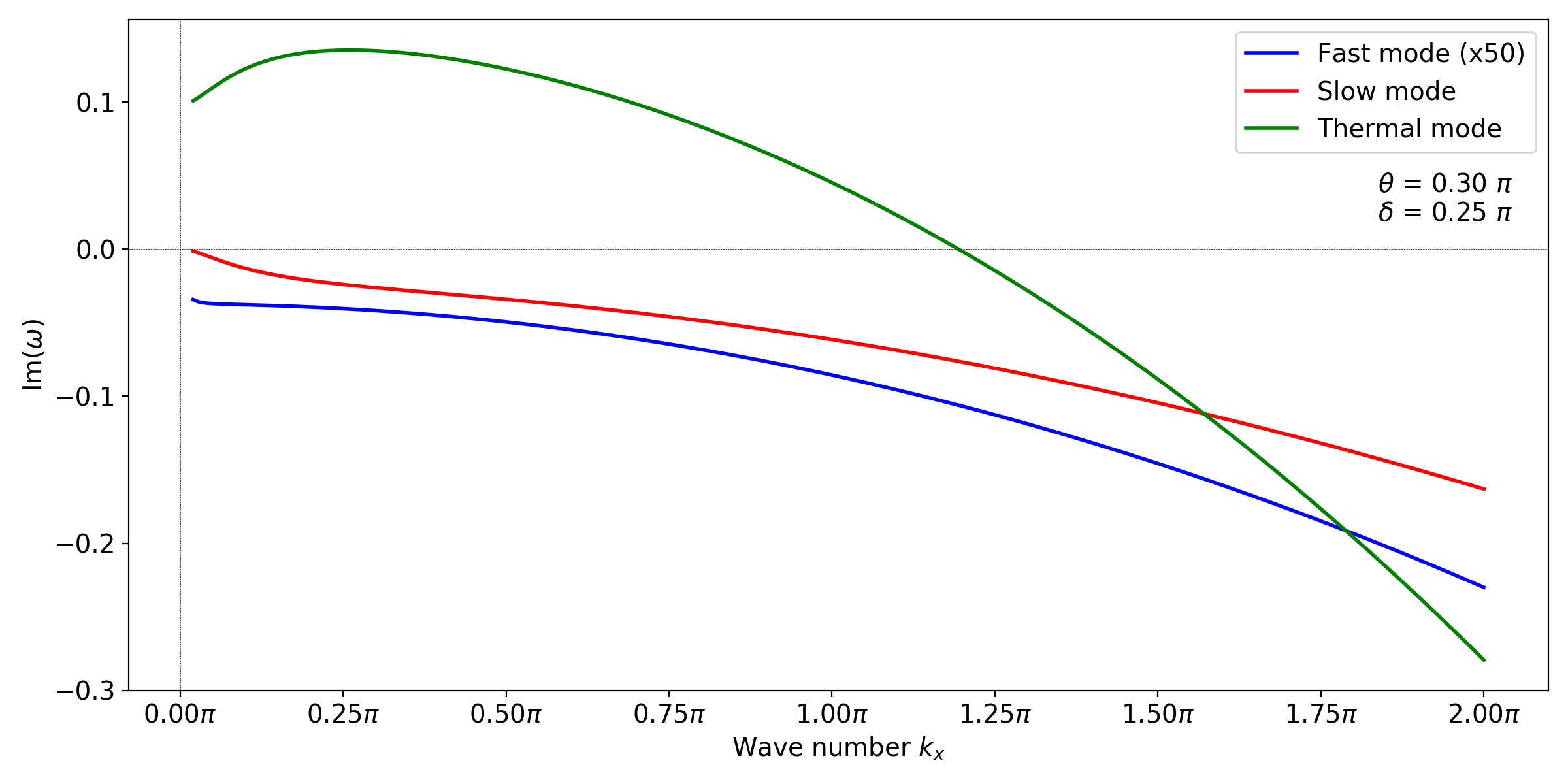}
        \caption{Growth rates of the MHD modes as a function of wave number, for $\bk$ aligned with the $x$-axis for an angle $\theta = 0.3\pi$ and $\delta = \pi/4$.
                         The thermal mode is stable above $k_x \approx 1.20\pi$, which is the critical wave number for these conditions. 
                         The growth rate of the fast mode was multiplied by $50$ for clarity. The equilibrium temperature is $0.50$ MK; the density and magnetic field are the default values.}
        \label{fig: w_vs_k}
\end{figure}
\begin{figure}[h]
        \centering
        \includegraphics[width=\columnwidth]{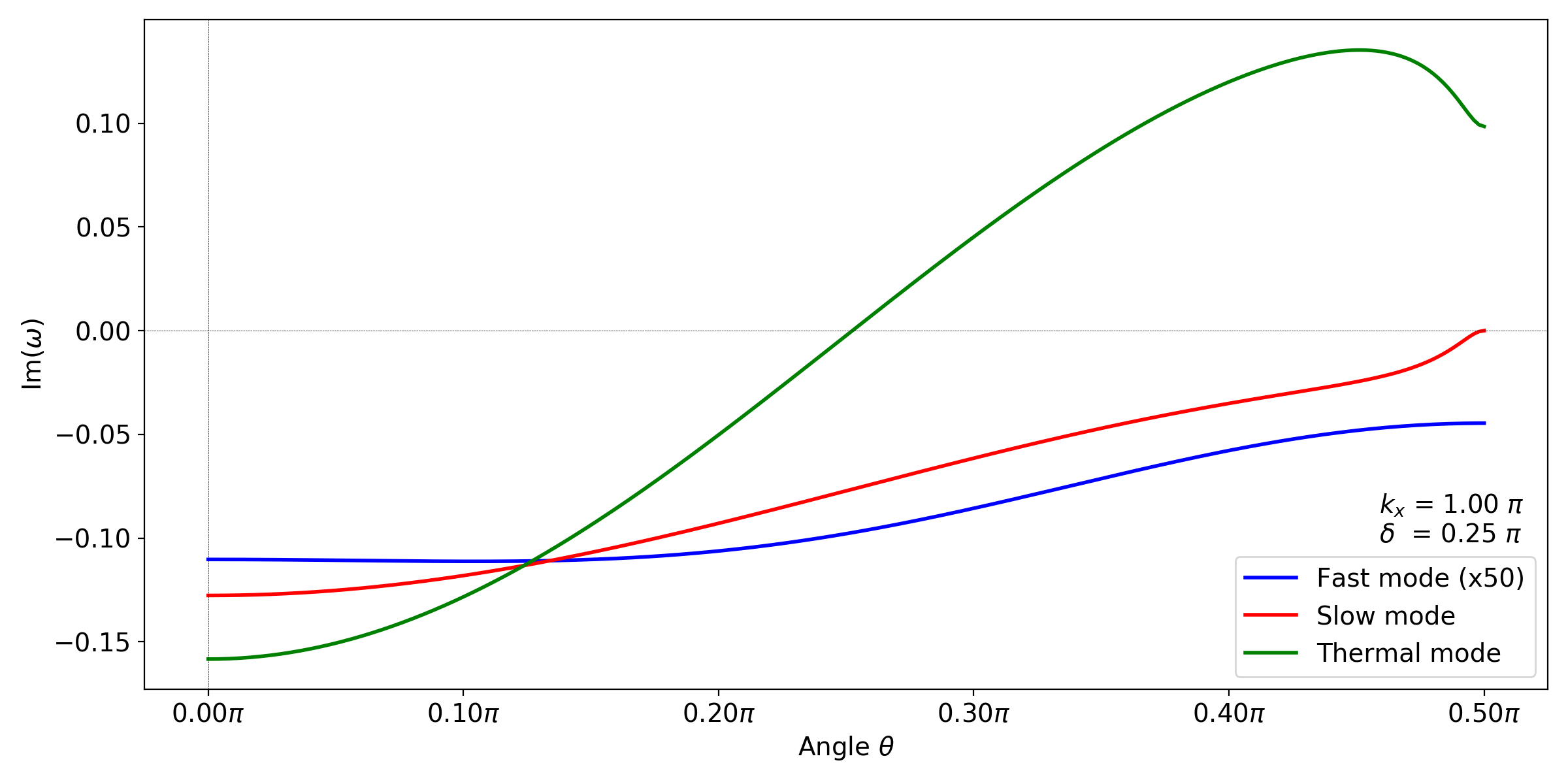}
        \caption{Growth rates of the MHD modes as a function of $\theta$ for a fixed value of $\delta = \pi/4$ and $k_x = \pi$. The wave vector $\bk$ is aligned with the $x$-axis. 
                                  The thermal mode becomes unstable for $\theta > \pi/4$, at this point parallel thermal conduction is too ineffective for stabilisation due to the alignment between $\bb_0$ and $\bk$. The growth rate of the fast mode was multiplied by $50$ for clarity.
                                  The equilibrium temperature is $0.50$ MK; the density and magnetic field are the default values.}
        \label{fig: w_vs_theta}
\end{figure}

Figure \ref{fig: w_vs_k} shows the growth rate of all three relevant modes under typical solar coronal conditions, where we now fix the temperature to 0.5 MK and a field orientation. The growth rates are shown as a function of wave number. The wave vector $\bk$ was taken along the $x$-axis. The slow and fast wave modes are damped everywhere under these conditions, but the thermal mode shifts from stability to instability when going below the critical wave number. The angles $\theta$ and $\delta$ were given a value of $0.30\pi$ and $0.25\pi$, respectively. 
We note that all growth rates in this work are specified in code units, where a unit of time is equal to $t_n \approx85.87$ seconds.

Figure \ref{fig: w_vs_theta} on the other hand indicates that solely considering the critical wave number is not sufficient for instability of the thermal mode. The wave number was taken to be $k_x = \pi$, which according to Fig. \ref{fig: w_vs_k} should indicate instability. However, now we vary the angle $\theta$, and when it becomes too small (for fixed $\delta = \pi/4$) the magnetic field becomes too aligned with the wave vector such that the contribution of parallel thermal conduction is particularly effective in smoothing out emerging temperature variations. Only for values of $\theta > \pi/4$ thermal conduction parallel to the field lines becomes inefficient enough, and consequently enables the thermal mode to become unstable for these conditions. Armed with this knowledge, all simulations performed further on achieve conditions where the slow modes are damped, but the thermal mode is inherently unstable.

\subsection{Thermal instability onset in 2D}    \label{subsect: 2D}
First we consider a purely 2D set-up where we superimpose two slow modes, one along each axis. This is done by calculating the eigenvalues and eigenfunctions for each $\bk$ and exciting them in a similar way as described before.
These waves interact with each other, but are both damped owing to the slow-mode stability shown in Figures \ref{fig: w_vs_k} and \ref{fig: w_vs_theta} (the growth rates $\omega_I$ are negative). Because we excited the eigenfunctions in a regime
that is unstable to the thermal mode, this regime eventually becomes excited and takes over as soon as the waves are sufficiently damped out. The thermal mode starts to grow, locally increasing the density, leading to an increased cooling contribution and a lower temperature. This eventually leads to high-density, low-temperature filaments, which break up and redistribute themselves across the domain \citep{claes2019}. It is important to stress that the term `filament' in this context does not specifically refer to the solar feature, but rather to a simple high-density, low-temperature region.

\begin{figure}
        \centering
        \includegraphics[width=\columnwidth]{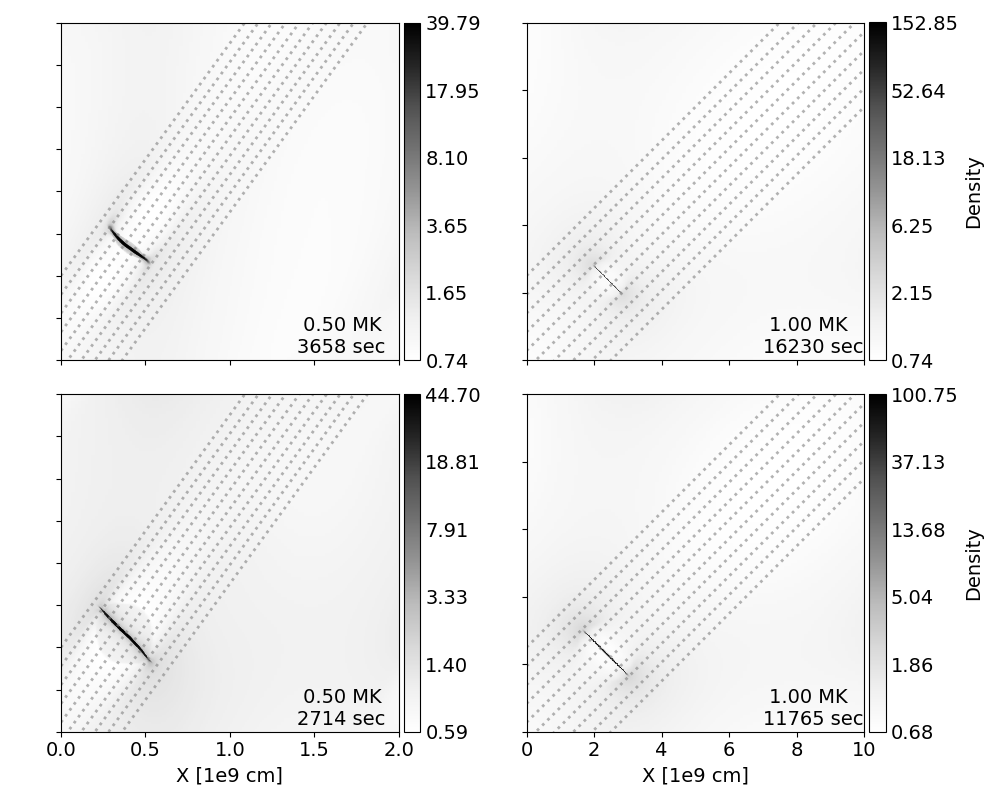}
        \caption{Density view at the onset of TI for a background temperature of $0.50$ MK (left column) and $1$ MK (right column), with conduction (top rows) and without conduction (bottom rows).
                                  The formation of the high-density filament happens approximately perpendicular to the magnetic field (dotted lines) in all cases. The density is normalised to $\rho_n$.}
        \label{fig: 2d_instability_onset}
\end{figure}

Figure \ref{fig: 2d_instability_onset} shows the onset of TI, just after the high-density filament is formed but before the breaking-up phase.
Two equilibrium temperatures are considered, $0.50$ MK in the left column and $1$ MK in the right column, thermal conduction is included for the top row and omitted for the bottom row. Because of the stabilising effect of thermal conduction, the onset of TI occurs at a later time than without conduction (the growth rate is smaller); the actual physical times are denoted on the figures for each case.
The magnetic field lines are superimposed using dashed lines on each figure, and make an angle $\theta = 0.30\pi$ and $\theta = 0.25\pi$ with the $x$-axis for the left and right columns, respectively.
We note that the right two panels have a larger domain size because the thermal mode is less unstable at one million Kelvin. We thus needed to adjust the box size to ensure that the wavelength of the system becomes larger and $|\bk|$ smaller, satisfying the critical wave number criterion for instability.

\subsection{Ram pressure and filament fragmentation}    \label{subsect: ram_pressure}
The onset of filament formation depicted in Fig. \ref{fig: 2d_instability_onset} indicates that in all cases the high-density filament seems to form nearly perpendicular to the magnetic field, such that material is literally flowing towards the high-density region along the field lines. This effect in turn creates a velocity difference on opposite sides of the filament, consisting of two counter-streaming flows of matter and a high-density region in between. 

In order to investigate this effect in greater detail we chose one of the simulations shown above, namely the $0.50$ MK case with conduction, increased the number of AMR levels to 7 and let it run far into the non-linear regime, long after the filament has formed.
This increase in refinement achieves a maximum effective resolution of $6400 \times 6400$, corresponding to a smallest cell size of $3.125$ km. Since the typical length scale of coronal rain is of the order of a few $100$ kms, any occurrence of small high-density blobs in the far non-linear regime are resolved. All figures in this subsection refer to this simulation.

\begin{figure}
        \centering 
        \includegraphics[width=\columnwidth]{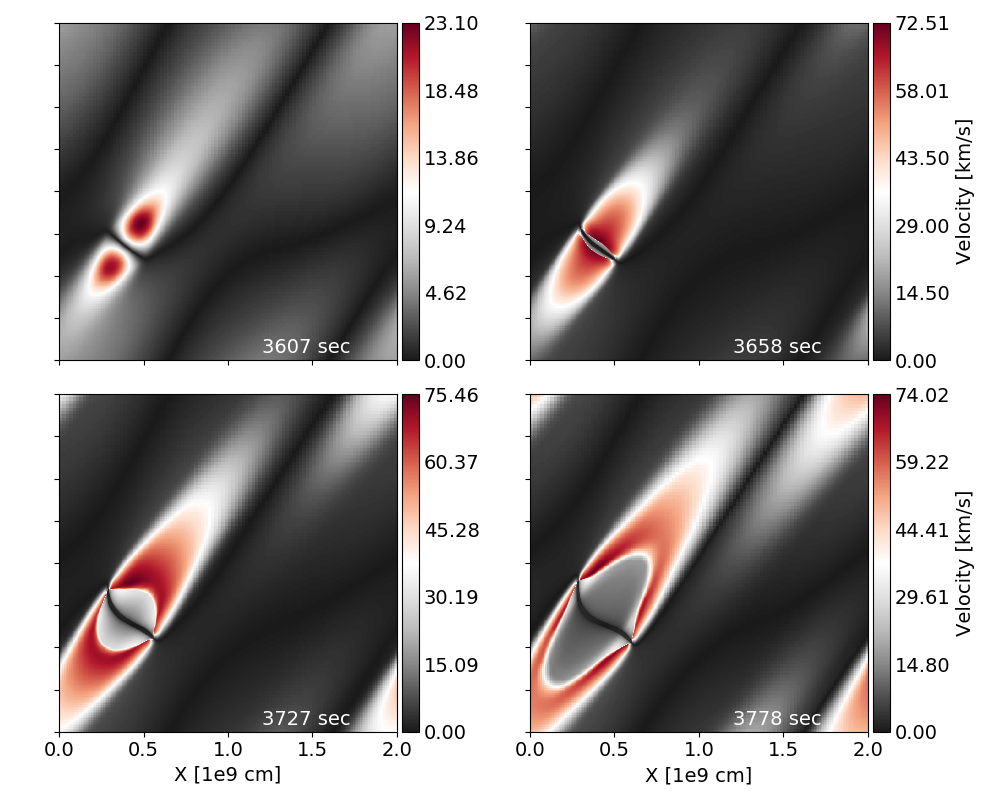}
        \caption{Plots of the velocity magnitude for the simulation corresponding to the top left panel of Fig. \ref{fig: 2d_instability_onset} (but with more AMR levels). 
                         The first collision of low-pressure induced inflows along the field lines produces a rebound slow shock on either side of the filament, travelling outwards.}
        \label{fig: 2d_velocity_plots}
\end{figure}

\begin{figure}
        \centering
        \includegraphics[width=\columnwidth]{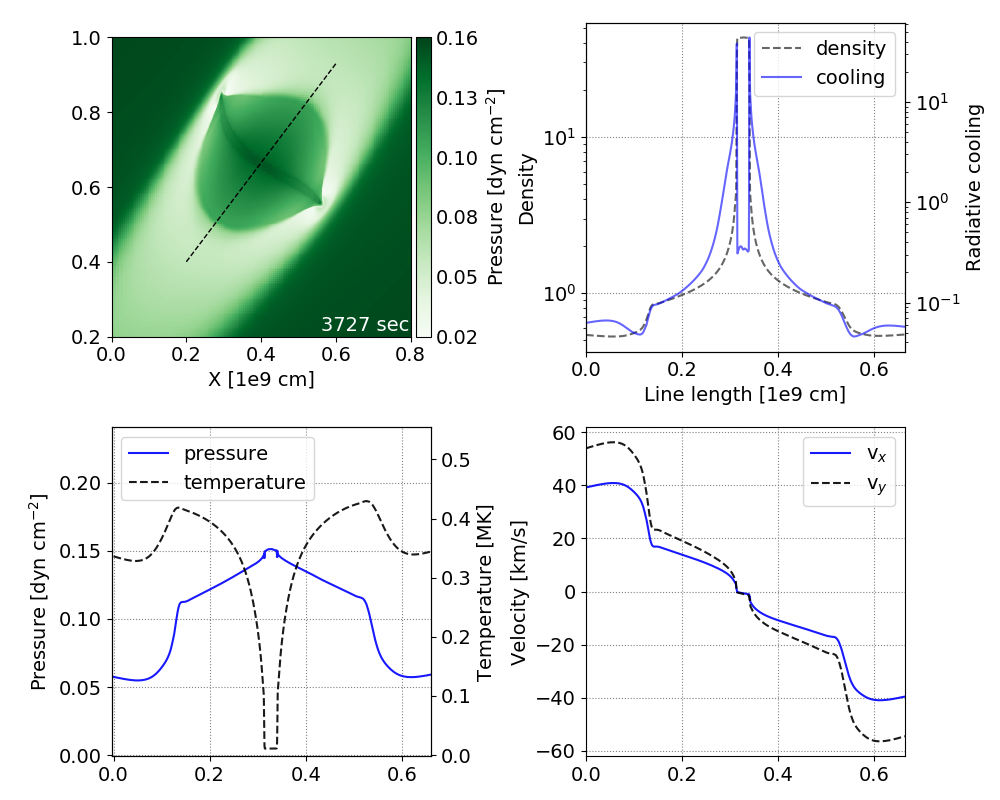}
        \caption{Pressure view (top left panel) along with various physical quantities along the selected line indicated, crossing the high-density blob. Density is normalised to $\rho_n$.}
        \label{fig: plotoverline}
\end{figure}

Fig. \ref{fig: 2d_velocity_plots} shows a detailed view of the formation process of the high-density region, at four different snapshots. The magnitude of the velocity field is shown, starting at the onset of filament formation. There is clearly a large velocity difference on either side of the high-density region owing to the infalling
matter on a trajectory aligned with the magnetic field. The filament itself appears mostly stationary, representing a real in situ formation process. It is enclosed by an outwards moving shock front on both sides. The slow and Alfv\'en wave velocities can be roughly estimated near to the shocked region, yielding $\approx 60-70$ km/s and $\approx 400$ km/s, respectively, such that we are dealing with slow MHD shocks enclosing the filament. These findings are in line with previous results obtained by \citet{xia2012} and \citet{fang2015shocks}, where this rebound shock phenomenon was also encountered in their simulations on prominence formation and coronal rain, respectively.
Fig. \ref{fig: plotoverline} shows a plot of the different physical quantities over a line spanning the high-density blob and the shock fronts. The top left panel shows a zoomed-in view of the pressure, with the selected line indicated in black. Density and radiative cooling jump about two orders of magnitude when crossing
the blob. Both gas pressure and temperature show a considerable increase over the shock fronts due to the compression and shock heating of the plasma, the blob itself is at a temperature of approximately $10000$ K. The bottom right panel shows the velocity components over the selected line, dropping almost to zero inside the blob, indicating a near stationary state. Sharp increases again denote the shock front itself, impacting the higher-velocity inflowing plasma.

The fourth panel of Fig. \ref{fig: 2d_velocity_plots} shows a later stage, in which the shocks have propagated even further and start to fan out. These rebound shocks are generated at the same time as the formation of the high-density region and are caused by
the first collision of low-pressure induced inflows of matter. After the filament has formed the inflows fall into the transition region between the condensations and shock fronts, without generating new shocks. This effect in turn feeds matter to the filament, causing it to extend further. The shock fronts sweep up and shock-heat matter when travelling outwards; their propagation speed is mostly determined by the gradually matter-evacuated environment of the filament. This implies faster shock propagation perpendicular to the filament (parallel to the field), and slower propagation perpendicular to the field lines (parallel to the filament). This in turn causes a pinching effect related to the ram pressure, as the ends of the filament are pinched between the outer edges of the rebound shocks. This is an inherently unstable configuration in which any small velocity difference on either side causes the filament to stretch out or undergo a torque. This effect is also clearly visible on the last panel of Figure \ref{fig: 2d_velocity_plots}.

\begin{figure}
        \centering
        \includegraphics[width=\columnwidth]{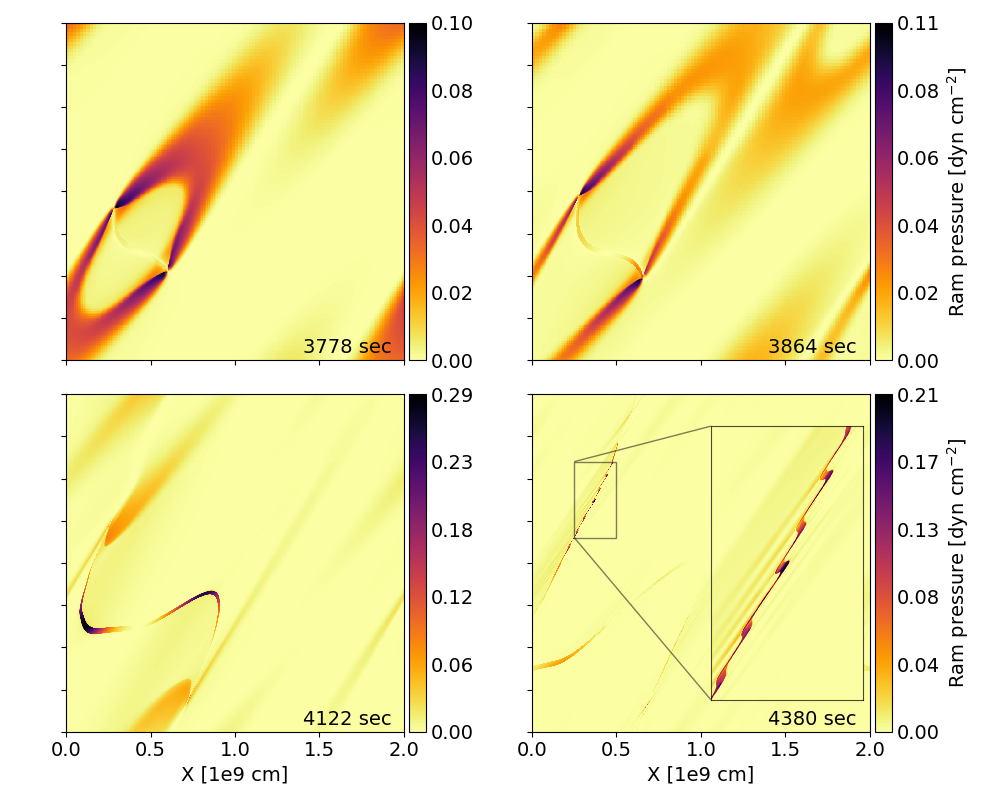}
        \caption{Ram pressure at four different snapshots after filament formation. The first panel corresponds to the fourth panel in Fig. \ref{fig: 2d_velocity_plots}. 
                         The sequence clearly shows how ram pressure is responsible for the stretching (top right and bottom panels) of the filament.
                         The inset on the fourth panel is a zoom-in on part of the filament, where small ram pressure imbalances cause instabilities and fragmentation through the thin-shell instability.}
        \label{fig: ram_pressure}
\end{figure}

The pinching phenomenon becomes very clear when looking at the ram pressure surrounding the system. Since we are dealing with two opposite velocity vectors, but parallel to the magnetic field, we can assume that the original inflow is perpendicular to the filament surface (and later to the shock fronts). In this case the expression for the ram pressure simplifies to
 \begin{equation}
        P_\text{ram} = \rho |\bv|^2,
 \end{equation}
 which is simply the density multiplied by the square of the velocity magnitude. Figure \ref{fig: ram_pressure} shows the ram pressure at four different snapshots, with the first one at the same time as the fourth panel of Fig. \ref{fig: 2d_velocity_plots}. Here the pinching effect is clearly visible, but, more importantly,
 there is an asymmetry present between ram pressures on either side of the filament ends. This becomes even more clear on the second panel in which the ram pressure on the bottom side of the right end is much higher than on the opposite side, and the same (but in reverse) holds for the left side of the filament.
 This effect stretches out the filament and redistributes it across the domain. However, at this point the original filament is thinned out by all the stretching, such that certain regions are particularly susceptible to small imbalances in velocity or pressure. This is exactly what is happening on the fourth panel of Fig. 
 \ref{fig: ram_pressure}, which shows a zoomed-in view of a part of the filament, indicated by the enclosing box. Small blobs start to form, driven by small imbalances in ram pressure, which cause fragmentation of the filament.
 This kind of phenomenon is typical for a thin-shell instability, first described in \citet{vishniac1983}.
 
 \begin{figure}
        \centering 
        \includegraphics[width=\columnwidth]{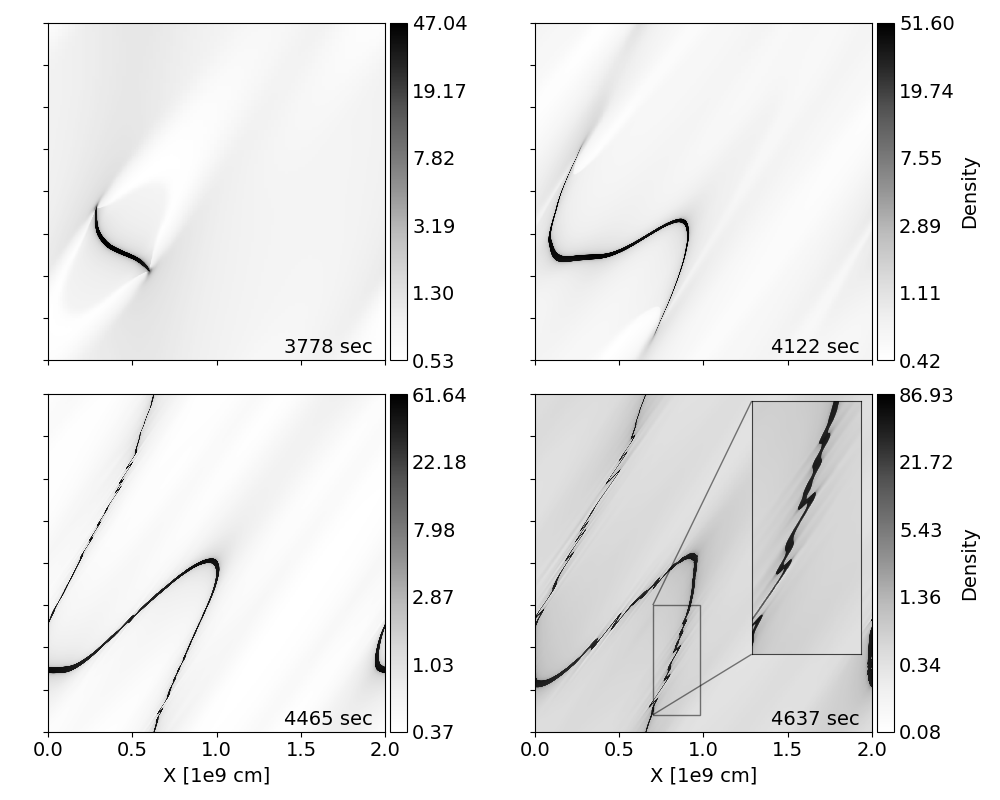}
        \caption{Density plots right after filament formation (first panel) and far into the non-linear regime (other panels). The filament starts to fragment on the third panel through the thin-shell instability.
                         Fragmentation has progressed even further in the last panel, where the inset zooms in on one of the regions that are breaking up into smaller blobs. Density is normalised to $\rho_n$.}
        \label{fig: density_plots}
 \end{figure}

Figure \ref{fig: density_plots} shows the density evolution right after filament formation, all the way up to the fragmentation process. The rebound shocks are visible on the first panel, the second panel shows how the filament is stretched out owing to ram pressure differences. Fragmentation starts to occur on the third panel near
to the outer ends of the filament through the thin-shell instability. In the fourth panel there is even more fragmentation, where now even the central filament region is breaking up.
The inset zooms in on one of the fragmenting regions, the smallest blobs have a width of the order of a few tens of grid cells (at the highest resolution), corresponding to physical length scales of a few 100 km, consistent with typical coronal rain sizes.

\subsection{Thermal instabilities in 3D}        \label{subsect: 3D}
Next we look at how the above translates to a 3D set-up in which three slow waves are superimposed, again taken along the coordinate axes. All slow modes are damped but unstable to the thermal mode, hence the entire process towards TI are similar as in 2D: the waves interact with each other and dampen out, but owing to the TI a runaway cooling reaction is triggered and a high-density region forms.

\begin{figure}
        \centering
        \includegraphics[width=\columnwidth]{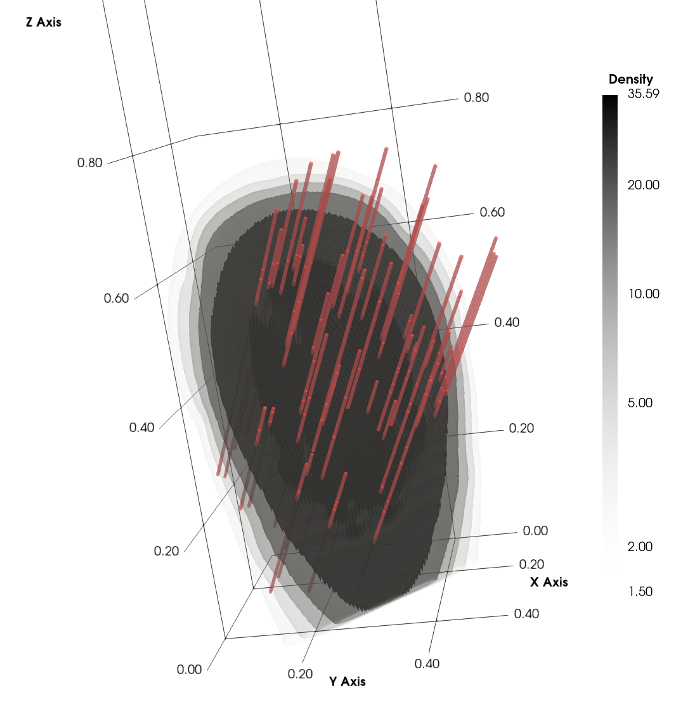}
        \caption{Density isocontours at the onset of filament formation in three dimensions. Similar to the 2D case formation of the high-density region appears to happen nearly perpendicular to the magnetic field lines, shown in red.
                         Density is normalised to $\rho_n$, the snapshot is taken at $t = 2920$ s. This is the onset of TI; no fine structure is yet present.}
        \label{fig: 3d_onset}
\end{figure}

Figure \ref{fig: 3d_onset} shows the onset of the high-density region in three dimensions for the same equilibrium parameters as the $0.50$ MK simulation in 2D (Figs. \ref{fig: 2d_velocity_plots} - \ref{fig: density_plots}). The only exception is that in this case the magnetic field has a different alignment owing to the 3D set-up, making an angle $\theta = 0.30\pi$ and $\delta = 0.25\pi$.  Density isocontours are shown, and similar to the 2D case the formation of the high-density region appears to occur almost perpendicular to the magnetic field lines, which are drawn in red on the figure. 

\begin{figure}
        \centering
        \includegraphics[width=\columnwidth]{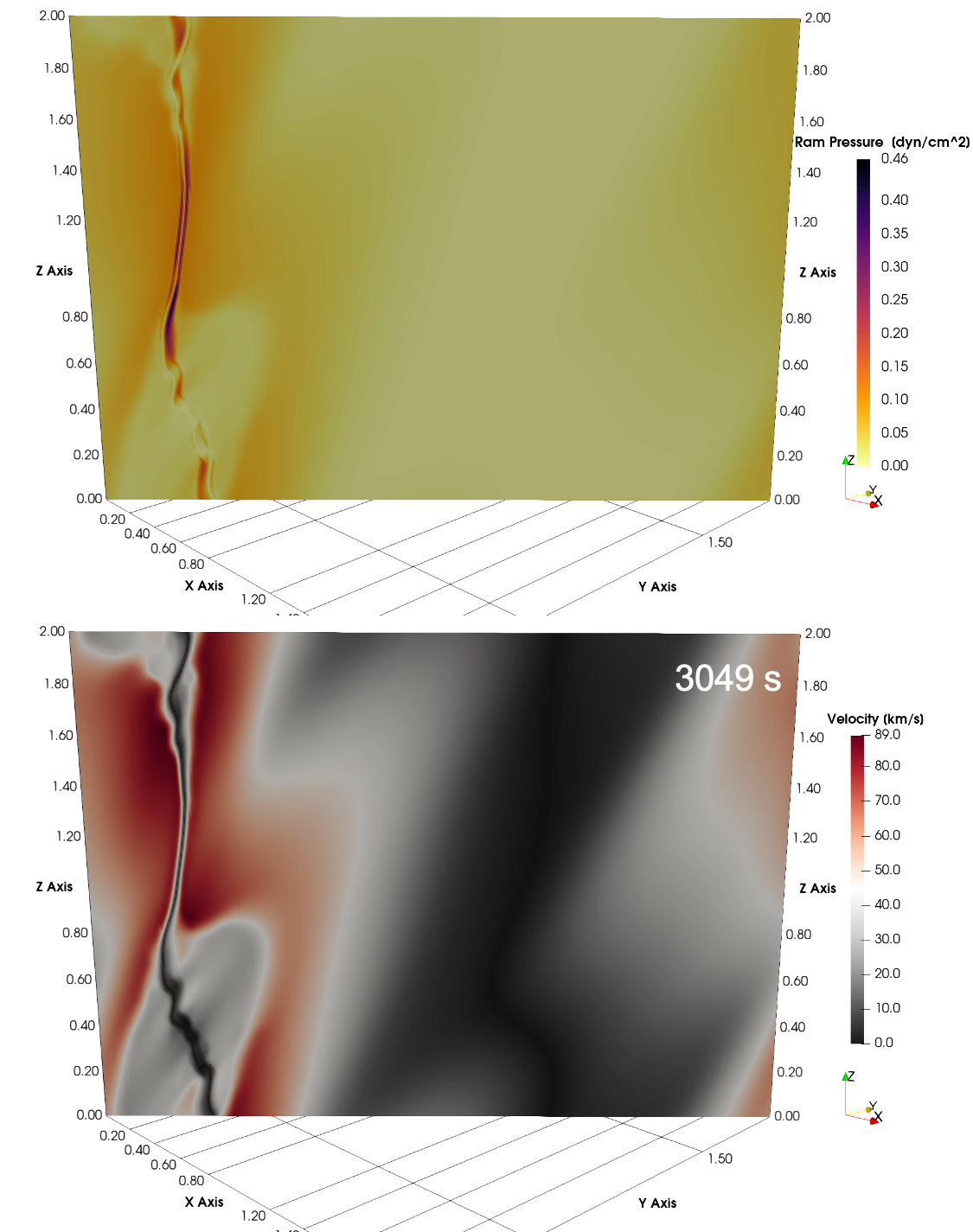}
        \caption{Slices showing ram pressure slice (top panel) and velocity magnitude (bottom panel) at $\approx$ 0.85 h. Two outwards travelling rebound shocks are present which fan out over time. 
                         Kinks in the filament and ram pressure differences on opposite sides denote the onset of thin-shell instability, fragmenting the high-density region.}
        \label{fig: pv_snapshots}
\end{figure}

Figure \ref{fig: pv_snapshots} depicts a slice diagonally taken across the 3D domain, parallel to the $z$-axis, showing the ram pressure distribution and velocity magnitude.
Various regions of higher ram pressure surround the filament, alternated by lower ram pressure voids. Similar to the 2D case, outwards travelling rebound shocks can be seen that sweep up and shock heat the plasma.
The actual pressure and velocity values in these regions are comparable to the 2D results and are again slow wave shocks. The pinching effect caused by the expanding shock fronts is again present, albeit less pronounced than for the 2D case discussed before.

\begin{figure}
        \centering
        \includegraphics[width=\columnwidth]{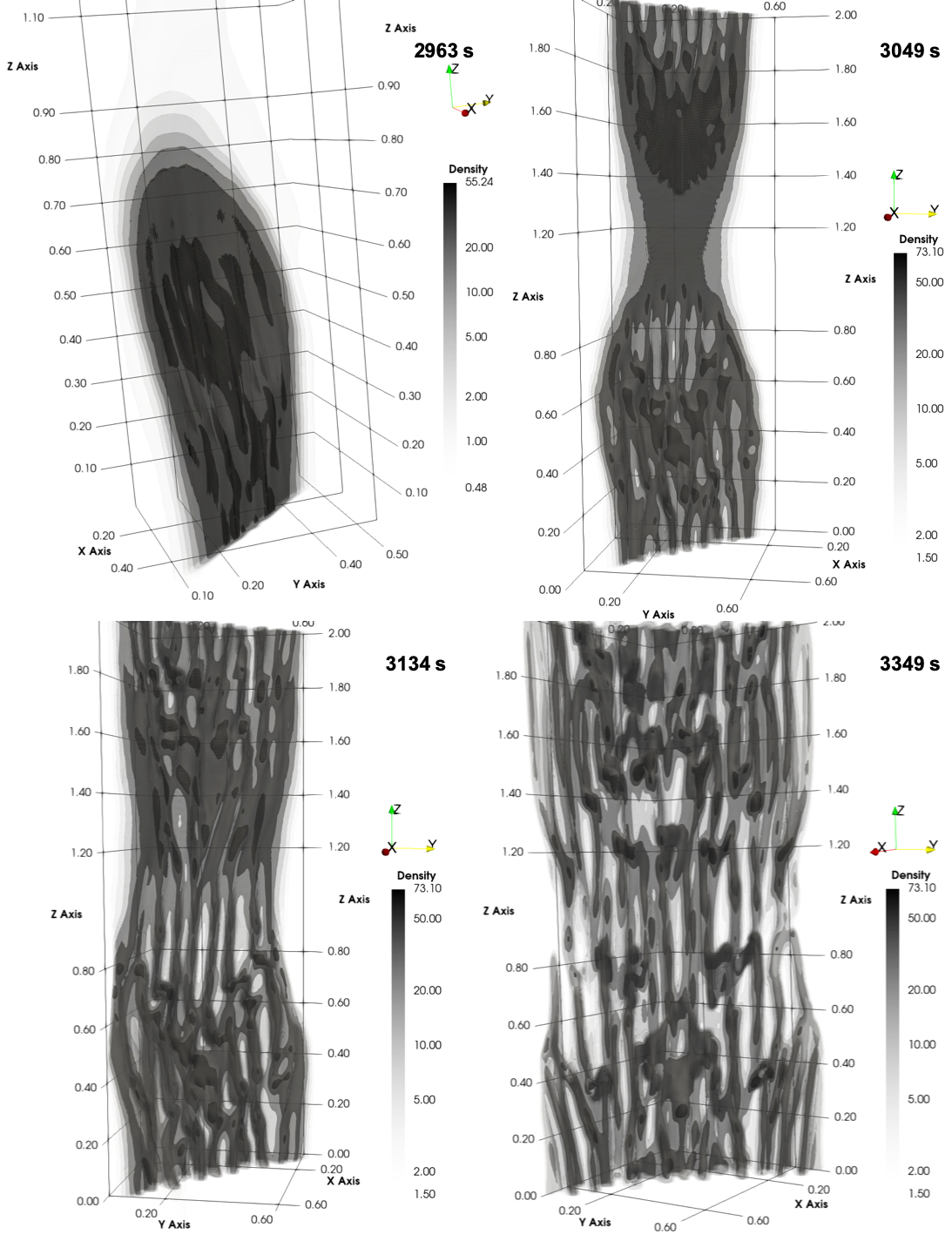}
        \caption{Density isocontours at four different snapshots in the 3D simulation, the physical time is denoted on each panel. The filament has fragmented and appears to coalesce into multiple dense blobs,                          interconnected by various higher-density strands, forming a morphologically complex web of fine structure.}
        \label{fig: density_overview}
\end{figure}

Figure \ref{fig: density_overview}  shows density isocontours of different snapshots, where the top right panel corresponds to the same time as the diagonal slices in Fig. \ref{fig: pv_snapshots}. Two major differences between the 2D and 3D cases should be noted. First, because of periodic boundary conditions the bottom part of the filament re-enters the domain at the top as shown in the first two panels of Fig. \ref{fig: density_overview}. However, this has no major influence on the substructure itself, as fragmentation and the formation of elongated strands occurs before this happens. The first panel of Fig. \ref{fig: density_overview} already shows slightly higher-density regions and some weak filamentary substructure which become even more distinct on the second panel, and by the time the third panel is reached the process results in one large high-density region spanning the entire $z$-axis of the domain, showing a very complex morphology. On the last panel, the higher-density regions have coalesced into blobs, interconnected by a web of slightly less dense, elongated strands.

\begin{figure}
        \centering
        \includegraphics[width=\columnwidth]{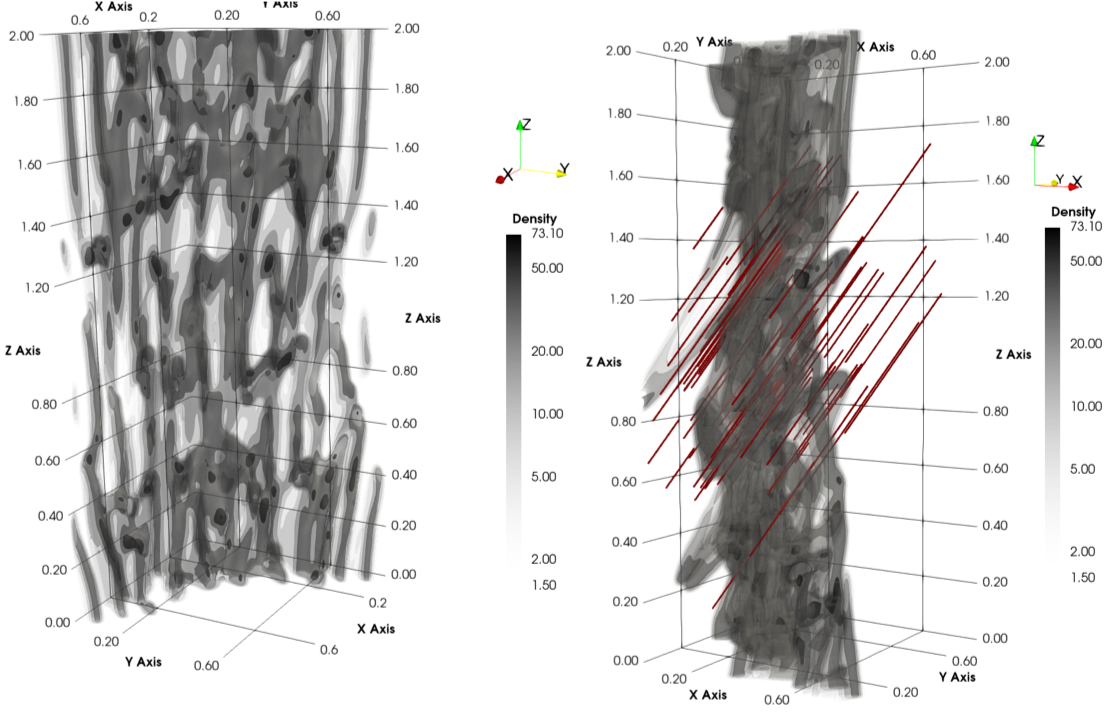}
        \caption{Density isocontours at time $t = 3565$ s ($\approx$ 0.99 h). The left and right panels correspond to different viewing angles, the magnetic field lines are denoted in red.}
        \label{fig: density_last_snapshot}
\end{figure}

The second difference to discuss is the alignment between the high-density threads and the background magnetic field. Figure \ref{fig: density_last_snapshot} shows the final snapshot of the 3D simulation for two different viewing angles, with the right panel rotated about 90 degrees counterclockwise around the $z$-axis with respect to the left panel, and the magnetic field lines superimposed in red. On the left the overall picture is not so different from the last panel in Fig. \ref{fig: density_overview}, except that some long strands have broken up into multiple, smaller `blob'-like pieces with high density. What should be noted however is that these strands are not at all aligned with the field lines. We see that some blobs appear to move outwards, away from the main high-density region, guided by the magnetic field. This implies that the actual condensation is aware of the magnetic field orientation, but is not necessarily completely aligned with it.

\begin{figure}
        \centering
        \includegraphics[width=\columnwidth]{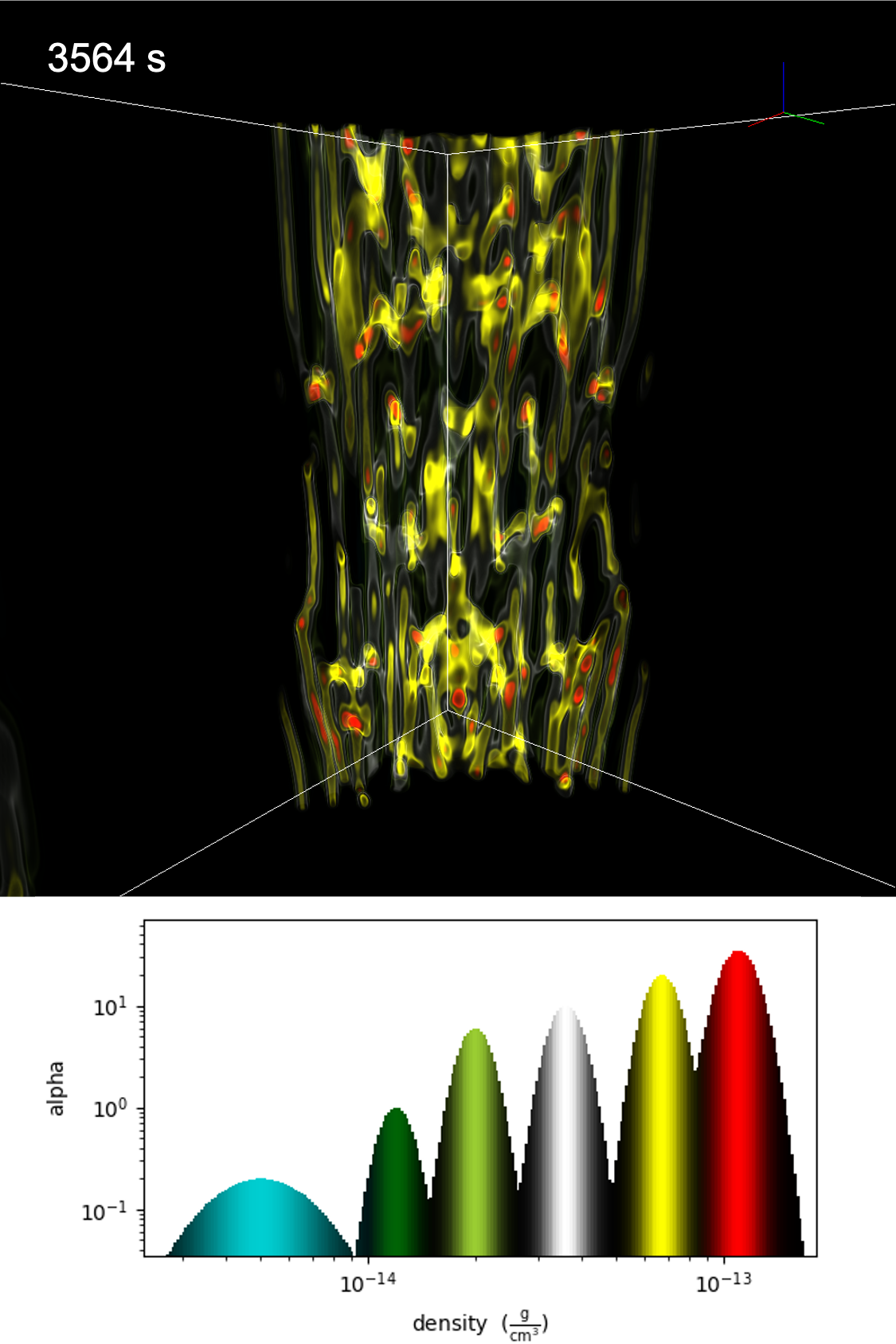}
        \caption{Volume rendering of the filamentary fine structure at $t \approx$ 0.99 h (last snapshot). The transfer function is customised and tweaked to highlight high-density regions and is shown on the bottom panel.
                         The solid white lines denote the domain boundaries. This rendering was made by \texttt{yt} \citep{YTpaper} using the newly developed AMRVAC frontend. (An animation of this figure is available in the on-line journal.)}
        \label{fig: volume_renderings}
\end{figure}

This last conclusion becomes very clear when looking at a volume rendering of the high-density filaments in Figure \ref{fig: volume_renderings}. The transfer function was customised and tweaked for high density such that dense regions appear orange-red, while less dense regions have a more green-blue colour. The transfer function itself is shown on the bottom panel, where the alpha value is shown as a function of density. 
The red and yellow Gaussians are centred at $1.1 \times 10^{-13}$ and $6.7 \times 10^{-14} $g cm$^{-3}$, respectively, with high alpha values to make these regions stand out. 
The Gaussians denoting the light green and white colours are centred at $2 \times 10^{-14}$ and $3.6 \times 10^{-14}$ g cm$^{-3}$, respectively, with an approximately similar alpha value.
The darker green and blueish parts of the transfer function have their centre at $1.2 \times 10^{-14}$ and $5 \times 10^{-15}$ g cm$^{-3}$, respectively, where the alpha values were significantly lowered as to reduce the contribution of the ambient medium, which would otherwise completely blur the image. In this rendering the multi-stranded nature of the main high-density regions becomes very clear, where the elongated threads are shown in a yellowish colour embedded in slightly less dense  strands, with the denser, cool blobs appearing red. The final snapshot of the simulation is shown, where we see some blobs moving away from the high-density region in a direction that appears to be guided by the magnetic field lines. An animated view of this figure is available in the on-line journal.

To stress the emerging fine structure and field misalignment even further, we created synthetic H$\alpha$ views of the 3D simulation discussed above using a method described in \citet{heinzel2015}. This method allows us to calculate the ionisation degree and electron density from local pressure and temperature variables in the grid, which in turn gives an indication of the absorption coefficient $\kappa_\nu$ for the H$\alpha$ line. This is used to estimate the optical thickness, which can then be integrated along the line of sight to obtain the specific intensity in H$\alpha$. A short description of this method is given in Appendix A. Figure \ref{fig: synthetic} shows synthetic H$\alpha$ views obtained by integration along the three coordinate axes for three different snapshots.  Especially on the third column of panels, corresponding to the last snapshot in the simulation and in Fig. \ref{fig: volume_renderings}, the complex morphology of the filament threads and high-density blobs is clearly visible. The line-of-sight integration along the $z$-axis clearly shows the magnetic field awareness, aligned diagonally from the bottom left to the top right corner when projected onto the $xy$-plane ($\delta = 0.25\pi$).

\begin{figure}
        \centering
        \includegraphics[width=\columnwidth]{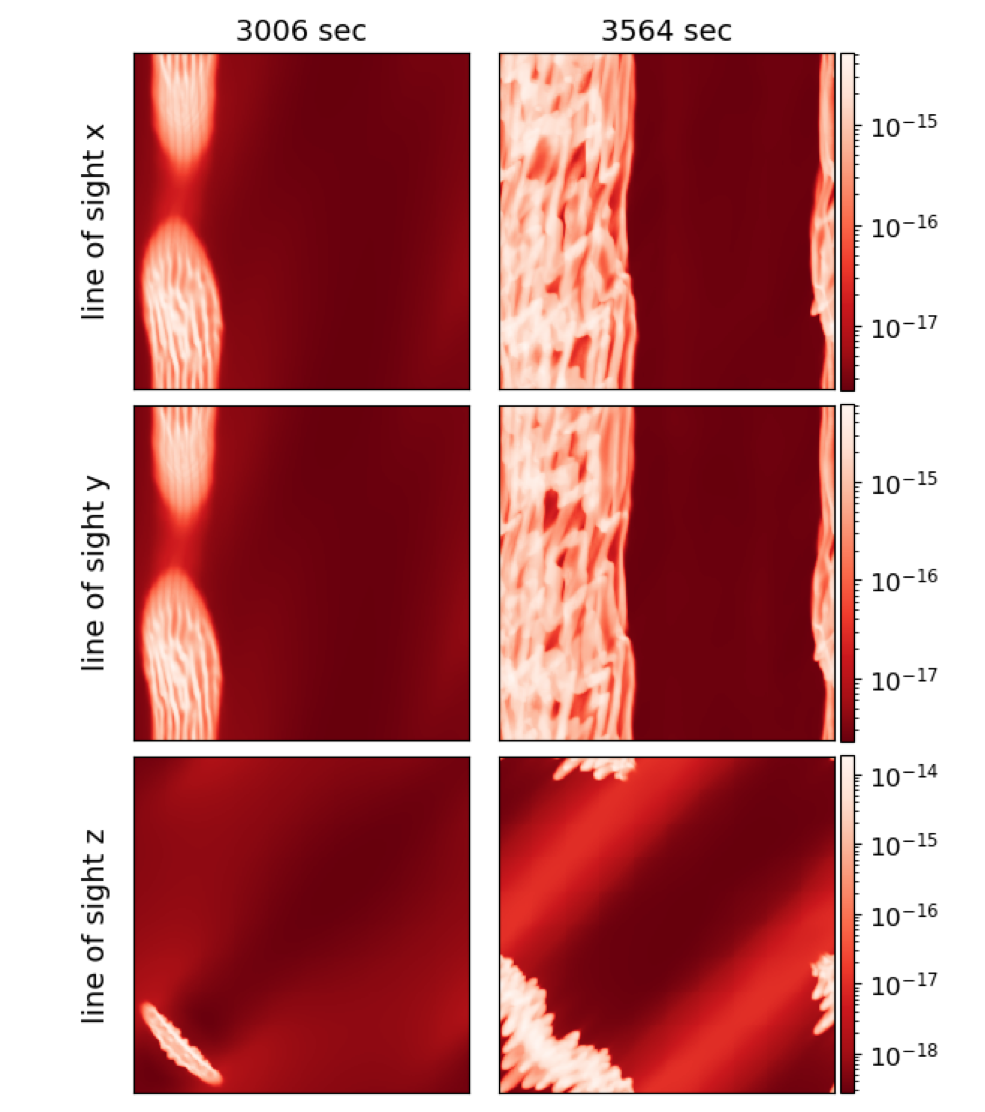}
        \caption{Synthetic H$\alpha$ views from the 3D simulation in Fig. \ref{fig: density_overview} for line-of-sight integrations along the tree coordinate axes (rows) for two different snapshots (columns). 
                         The resulting projection further stresses the complex morphology of emerging fine structure; the bottom row of panels clearly shows the awareness of $\bb_0$, aligned diagonally when projected onto the $xy$-plane.}
        \label{fig: synthetic}
\end{figure}

The actual implications of these last findings are hence twofold. Firstly, the fact that fine structure can be aligned quite differently with respect to the magnetic field orientation directly extends to solar prominences and prominence substructure, 
implying that direct observation of fine structure alignment does not necessarily constrain the direction of the magnetic field.
Secondly, these simulations were done using a uniform equilibrium state, but can resemble local solar coronal conditions. This is especially relevant for the multi-stranded shape of solar prominences, since in the simulations discussed in this work these features seem to occur in a natural way through a combination of TI, thermal conduction effects, rebound shocks, and ram pressure differences. These latter two give rise to thin-shell instability; a pinching effect is caused by the outwards propagating shock fronts which elongates, turns and fragments the filament (2D), or through an early occurrence of ram pressure imbalances, immediately fragmenting the high-density region before it can undergo a torque (3D). In both cases slow rebound shocks fan outwards, after which we eventually see fragmentation of the filament with high-density blobs guided by the magnetic field lines and fine structure not at all aligned with the background magnetic field.

\section{Conclusions}
It has been widely assumed that the orientation of filament fine structure can give some indirect information on the magnetic field topology in solar prominences. To that end, we performed numerical 2D and 3D simulations of interacting slow waves with the inclusion of radiative cooling effects and anisotropical thermal conduction. Typical values for the density in our simulations were taken to be $2.34 \times 10^{-15}$ g cm$^{-3}$ with a magnetic field strength of 10 gauss.
Combined with an equilibrium temperature between 0.50 MK and 1 MK, this yields a plasma beta between 0.04 and 0.08. The main findings can be summarised in a few key points:
\begin{enumerate}
        \item In order for the TI to trigger, the local plasma conditions must correspond to a regime unstable to the thermal mode. Both slow and fast waves are damped for solar coronal conditions and in order for the thermal mode to be unstable in the presence of anisotropic
                           thermal conduction the critical wavelength has to be satisfied. In any situation in which that is not the case, all modes dampen out.
        \item When TI occurs the high-density region appears to form orthogonal to the magnetic field lines, in both 2D and 3D. Material flows towards the filament following the magnetic field lines, guided by the strong pressure gradient.
                           The first collision of low-pressure induced inflows creates rebound shocks that travel outwards and can be classified as slow MHD shocks.
        \item After formation of the condensation and shock fronts, matter passes through the transition region between the high-density filament and the shock fronts. This does not generate new shocks, but feeds the filament causing it to extend further.
        \item The shock fronts start to fan out, propagating faster in a direction perpendicular to the filament owing to a gradually matter-evacuated filament environment.
              As seen in the 2D simulations this creates a pinching effect, eventually stretching out the filament and exerting a torque. However, this stretching leads to thin sheets, which
                  are susceptible to the thin-shell instability through the process of ram pressure imbalances. This effect eventually fragments the filament, creating multiple high-density blobs with a physical length scale of a few 100 kilometres.
        \item These rebound shocks occur again in 3D, however, this time the thin-shell instability kicks in almost immediately, fragmenting the filament shortly after formation. 
                  The initial 3D pancake shape orthogonal to the magnetic field lines breaks up and starts to coalesce into multiple high-density blobs interconnected by lower-density threads.
        \item The blobs start moving in the direction of the background magnetic field, while the thread-like structures are in various directions, not at all aligned with the field lines. This is confirmed by volume renderings and synthetic H$\alpha$ views.
\end{enumerate}

The last three points suggest that the actual shape of the condensation after the process of TI plays a fundamental role. If the resulting filament is too thick it first undergoes stretching and possibly a torque, all driven by ram pressure imbalances.
If the filament is thin enough on the other hand, the thin-shell instability can efficiently cause fragmentation early on, resulting in a completely different spatial structure.
If this is also the case in arcade-like or flux rope structures relevant for solar prominences, this effect could explain the multi-stranded shape of coronal rain, or the different alignment of threads observed in prominences.

We could also speculate concerning the relation between the wavelength of the initial slow wave perturbations and the spatial extent of the filament. As is clear from Figure \ref{fig: 2d_instability_onset} the actual `length' of the filament varies between the different panels. First of all the filaments for the cases without conduction appear slightly more extended perpendicular to the magnetic field lines with respect to the cases with conduction. Additionally, since we increased the size of the numerical domain for the simulations with an equilibrium temperature of one million Kelvin, this implies that the filament is larger for this case. A rough estimate can be made of both filament sizes, yielding a length of $\approx$ 4 Mm for the left panels
and $\approx$ 36 Mm for the right panels. Since the size of the domain is representative for the wavelength of the perturbation, this might imply a spatially larger filament when the wavelength of the perturbation increases, but is open to speculation.

We also wonder what causes the halted transverse extension of the condensed regions. It is not entirely clear what exactly is the determining factor to constrain the spatial extent of the filaments, however, this may be related to the gradually matter-depleted region surrounding the high-density filament. After formation of both the high-density region and rebound shocks, matter is still inflowing through the transition region and feeding the filament, causing it to extend further. This inflow of matter naturally diminishes in intensity the further the surrounding regions are depleted of plasma, while the shock fronts keep moving outwards. At some point this matter inflow may become inefficient compared to the above-stated pinching effect, such that the latter becomes dominant and starts to guide the dynamical evolution of the filament itself. At this point spatial extension of the filament could come to a stop, while ram pressure imbalances and thin-shell instabilities take over.

We did not include an arc-like magnetic field in our simulations nor gravitational effects. However, if this natural emergence of blobs and fine structure also persists in arcade-like configurations such as solar prominences, these high-density blobs should fall down along the magnetic field lines under the influence of gravity, towards the loop footpoints, or collect in dips that are pre-existing or formed by accumulation of mass. In our simulations, the blobs move away from the initial high-density region guided by the magnetic field. This is in accordance with results obtained by \citet{xia2017} in their simulations of coronal rain, where this type of phenomenon was found in realistic stratified environments. 

The study of TI is not only relevant in a solar context, but extends to almost every domain in astrophysics. It can for example be responsible for the formation and coalescence of gaseous clouds in parsec-scale environments, as shown by \citet{proga2015, waters2019}, and recent work even holds it responsible for clumpiness in active galactic nucleous outflows \citep{dannen2020}. The presence of TI on Galactic scales has also been hypothesised, where simulations done by \citet{peng2017} indicate the formation of dense molecular filaments through TI over a few hundred parsecs. Recent studies revisited the original work by \citet{field1965} using modern theoretical techniques, yielding new insights in the physics behind TI \citep{waters2019b, falle2020}. The wide applicability of TI to various astrophysical domains ensures this is a vary active research topic.

In future work, we plan to study non-linear TI development and its fine structure in gravitationally stratified slabs and actual coronal loop configurations. Ultimately, this will decide whether the misalignment effects found in this work persist in the large-scale structures we see as prominences in the solar corona.

\appendix

\section{Synthetic views}
We give a brief overview of the method described in \citet{heinzel2015} to create synthetic H$\alpha$ views based on local variables in the grid. First, the ionisation degree $\alpha_i = n_e / n_H$ and the parameter $f(T, p)$, which gives a relation between the electron density and hydrogen second-level population, are interpolated at each grid point from pre-computed tables in \citet{heinzel2015} using the local thermal pressure and temperature. The $H\alpha$ line absorption coefficient is then given by
\begin{equation}
        \kappa_\nu = \frac{\pi e^2}{m_e c}f_{23}n_2 \phi_\nu(\nu),
\end{equation}
where $m_e$, $c,$ and $n_2$ denote the electron mass, speed of light, and hydrogen second-level population, respectively. The latter can be found through
\begin{equation}
        n_2 = \frac{n_e^2}{f(T, p)}     \qquad \text{where} \qquad n_e = \frac{p}{\left(1 + \frac{1.1}{\alpha_i}\right)k_B T}.
\end{equation}
The H$\alpha$ line oscillator strength is given by $f_{23}$ and is equal to approximately 0.6407 \citep{f_halpha}. For the absorption line profile we adopted a Gaussian function $\phi_\nu(\nu)$ with a microturbulent velocity of 5 km s$^{-1}$. The optical thickness $\tau_\nu$ is obtained by integrating along the line of sight ($los$), that is,
\begin{equation}
        \tau_\nu = \int_{los} \kappa_\nu(\nu, l) dl.
\end{equation}
This is then plugged into the radiative transfer equation $I(\nu) = S(1 - \exp(-\tau_\nu))$ assuming a uniform source function $S$, which is taken to be equal to the diluted H$\alpha$ line-centre intensity at the solar disc centre, yielding the H$\alpha$ line intensity.

\bigskip\noindent
\footnotesize{Acknowledgements. Visualizations were performed using \texttt{Paraview} (\url{paraview.org}), 
\texttt{Python} (\url{python.org}) and \texttt{yt} (\url{yt-project.org}). The computational resources and services used in this work were provided by the VSC (Flemish Supercomputer Center), funded by the Research Foundation - Flanders (FWO) and the Flemish Government – department EWI. RK is supported by a joint FWO-NSFC grant G0E9619N and received funding from the European Research Council (ERC) under the European Unions Horizon 2020 research and innovation programme (grant agreement No. 833251 PROMINENT ERC-ADG 2018), and from Internal Funds KU Leuven, project C14/19/089 TRACESpace.}

\bibliographystyle{aa}
\bibliography{bibfile}

\end{document}